\documentclass[prl,twocolumn,showpacs,preprintnumbers,nofootinbib,amsmath,amssymb,nobalancelastpage]{revtex4}
\usepackage{graphicx}
\usepackage{bm}
\usepackage{dcolumn}
\usepackage{amsmath}
\usepackage{amssymb}
\usepackage{color}

\usepackage{aas_macros}

\newcommand{\sA}{\sigma}
\newcommand{\sigv}{\langle \sA v \rangle}

\newcommand{\degr}{^\circ}

\newcommand{\eps}{\epsilon}
\newcommand{\PSF}{{\rm PSF}}
\newcommand{\dNg}{dN_{\gamma}}
\newcommand{\Nnu}{N_\nu}
\newcommand{\Qi}{{Q_i}}
\newcommand{\F}{{\cal{F}}}

\begin{document}

\title{Dark matter line search using a joint analysis of dwarf galaxies with the Fermi Gamma-ray Space Telescope}
\author{Alex Geringer-Sameth}
\email{alex\_geringer-sameth@brown.edu}
\author{Savvas M. Koushiappas}
\email{koushiappas@brown.edu}
\affiliation{Department of Physics, Brown University, 182 Hope St., Providence, RI 02912}

\date{\today}

\begin{abstract}
We perform a joint analysis of dwarf galaxy data from the Fermi Gamma-ray Space Telescope in search of dark matter annihilation into a gamma-ray line.
We employ a novel statistical method that takes into account the spatial and spectral information of individual photon events from a sample of seven dwarf galaxies. Dwarf galaxies show no evidence of a gamma-ray line between 10 GeV and 1 TeV. The subsequent upper limit on the annihilation cross section to a two-photon final state is $3.9^{+7.1}_{-3.7}\times 10^{-26} {\mathrm{cm^3s^{-1}}}$ at 130 GeV, where the errors reflect the systematic uncertainty in the distribution of dark matter within the dwarf galaxies. 

\end{abstract}


\maketitle

The search for dark matter annihilation directly into a photon final state is extremely important because the line emission occurs at an energy that corresponds to the mass of the dark matter particle (or thereabouts if the second particle is a heavy neutral particle) \cite{1989PhRvD..39.3549R,1989PhLB..225..372B,1989PhRvD..40.2549G,1994APh.....2..261B,1995PhRvD..51.3121J,1996PhR...267..195J,2010JCAP...04..004J}. In addition, line emission is free of background contamination as no known astrophysical process can result in line emission at the energies of interest (a few GeV up to 10's of TeV). 

Recently there have been claims of the presence of a gamma-ray line at $E_\gamma = 130~ {\mathrm{GeV}}$  \cite{2012arXiv1204.2797W,2012arXiv1205.1045T,2012arXiv1206.1616S}. These studies, based on 3.5 years of data from Fermi, find a line emission signature from the direction of the  Galactic center. The interpretation of these results as dark matter annihilating directly to a photon final state implies a cross section of $\langle \sigma v \rangle \approx [10^{-27} -10^{-26}] \,\,{\mathrm{cm^3/s}}$. It is important to emphasize that this annihilation cross section is much larger than what one would expect from second order diagrams that lead to a two-photon final state (or a single photon and a $Z$ gauge boson or $h$ --- for a summary see e.g., \cite{1996PhR...267..195J}). Several dark matter interpretations for the alleged line feature have been offered \cite{2012arXiv1205.1045T,2012arXiv1205.1520D,2012arXiv1205.2688C,2012arXiv1205.3276C,2012arXiv1205.4151K,2012arXiv1205.4675L,2012arXiv1205.4723R,2012arXiv1205.6811B} while other work raises doubts about the statistical significance of the line and its interpretation as dark matter \cite{2012arXiv1204.6047P,2012arXiv1205.4700B}. A recent search by the Fermi collaboration did not detect the presence of  line emission in the  Galactic halo (including the Galactic center) \cite{2012arXiv1205.2739F}.

The  Galactic center is clearly a place of interest when it comes to dark matter annihilation because of its large expected dark matter density \cite{1998APh.....9..137B}. As the annihilation rate is proportional to the square of the number density of dark matter particles, its high density, coupled with its proximity to Earth, makes the Galactic center an attractive target for the search for an annihilation signal (e.g. \cite{2011PhRvD..84l3005H}). 

In this short note we perform a search for dark matter  annihilation to a photon final state in Milky Way dwarf  galaxies using data from the Fermi Gamma-ray Space Telescope (Fermi). By virtue of their pristine dark matter environment (absence of high-energy baryonic processes) and high concentration of dark matter, dwarf  galaxies have been used to place the strongest bounds to-date on the s-wave annihilation cross section of dark matter \cite{2011PhRvL.107x1303G,2011PhRvL.107x1302A}. Given the paucity of background contamination along the lines of sight to the dwarf  galaxies, it is natural to consider what limits the dwarfs may place on the annihilation cross section of dark matter into photon final states. 

The approach we take is similar to \cite{2011PhRvL.107x1303G,GSK12}. We perform a line search by testing, at each line energy $E_\gamma$, the null hypothesis that the observed data was generated by background processes.
Each hypothesis test is based on a test statistic $T$, which can be an arbitrary function of the data; however, it is vital that the choice of test statistic be made without reference to the data actually measured in the direction of the dwarf galaxies.  

We choose a simple form for the test statistic that combines the photon information from each of the dwarfs. Each photon $i$ within a Region of Interest (ROI) of size $1\degr$ is assigned a weight $w$ based on which dwarf $\nu$ it came from, its energy $E$, and its angular separation $\theta$ from location of the dwarf. We denote this set of properties as $\Qi:\{\nu, E,\theta \}$.
The test statistic $T$ is the sum of the weights of the photons detected within the ROIs centered on each dwarf:
\begin{equation}
T = \sum\limits_{\nu} T_\nu \label{eqn:Tunbinned}
\end{equation}
where the single-dwarf test statistic $T_\nu$ is 
\begin{equation}
T_\nu  =\sum\limits_{i=1}^{\Nnu} w(\Qi).
\label{eqn:Tunbinned1}
\end{equation}
Here, $\Nnu$ is the number of photons detected within the ROI centered on dwarf $\nu$. The weights $w(\Qi)$ and the total number of photons $N_\nu$ from each dwarf are random variables\footnote{In previous work on the analysis of the continuum gamma-ray emission from a combination of dwarf galaxies \cite{2011PhRvL.107x1303G}, the weight of each photon was determined only by which dwarf it came from. The test statistic was therefore the weighted sum of the total number of photons collected from each dwarf.}. 

To calculate the statistics of $T$ it is useful to divide the parameter space of energy and angular separation (for each dwarf) into infinitesimal bins, each labeled by $Q = \{\nu, E, \theta\}$. The number of photons detected in each bin is a random variable $X_{Q}$. This total number of photons is the sum of two random variables: the number of photons from dark matter annihilation $S_Q$ and the number originating from background processes $B_Q$ (i.e., $X_Q = S_Q + B_Q$). It can be shown that the weight function that maximizes an expected signal to noise ratio for a line emission search is \cite{GSK12} 
\begin{equation}
w(Q) = \frac{s_Q}{b_Q},
\label{eqn:wsearch}
\end{equation}
where $s_Q$ and $b_Q$ are the expected number of signal and background counts in the parameter space bin $Q$. 

Note that while $s_Q$ and $b_Q$ are infinitesimal quantities (being proportional to the size of the infinitesimal $Q$ bin) their ratio is finite. In addition, $s_Q$  depends on signal characteristics in such a way that the expected signal in any infinitesimal $Q$ bin is directly proportional to the annihilation cross section. Therefore, changing the annihilation cross section will simply scale the test statistic by a constant factor and will not affect any statistical conclusions, i.e. this weight function is optimally powerful for any cross section.

The quantity $T_\nu$ (Eq.~\ref{eqn:Tunbinned1}) is the sum of two terms: the weights of photons from dark matter plus the weights of background photons. These terms are independent variables so the probability distribution function (PDF) for $T_\nu$ is the convolution of these individual PDFs. As in \cite{2011PhRvL.107x1303G,GSK12} we model the background processes using data from the region surrounding each dwarf galaxy. The fundamental assumption made is that the processes which give rise to the background nearby the dwarf also generate the background at the location of the dwarf. 

For each dwarf we find the PDF of $T_\nu$ due only to background processes by sampling the photons in the region within 15$\degr$ of the dwarf. Sources from the second Fermi LAT source catalog \cite{2012ApJS..199...31N} are masked with $0.8\degr$ masks (the 95\% containment angle for photons with energies greater than 10 GeV \cite{LATperformance}). The sampling is performed by randomly placing 1$\degr$ ROIs over the 15$\degr$ field of view (rejecting those ROIs which overlap with a masked source, the ROI centered on the dwarf, or the boundary of the field of view). The photons in these ROIs are then weighted according to  Eq.~\ref{eqn:wsearch} and summed as in Eqs.~\ref{eqn:Tunbinned} and \ref{eqn:Tunbinned1}.

In order to derive the PDF of $T$ due to an annihilation signal, consider first a single dwarf $T_\nu$ as given in Eq.~\ref{eqn:Tunbinned1}. The quantity $T_\nu$ is the sum of $N_\nu$ independent, positive random variables (the weights), where $N_\nu$ is drawn from a Poisson distribution with mean $\mu_\nu$, the expected number of dark matter photons from dwarf $\nu$. This distribution is known as a compound Poisson distribution \citep{adelson1966}. The PDF for $T$ due to dark matter (Eq.~\ref{eqn:Tunbinned}) for all the dwarfs is therefore the convolution of the individual compound Poisson distributions for each of the dwarfs.
The PDF for each weight in the sum $T_\nu$ is the same and is found by dividing the energy-angular separation plane into infinitesimal bins and computing the probability that a detected dark matter photon will land in each bin. The weight assigned to a photon landing in each bin is set by Eq.~\ref{eqn:wsearch}.

There are several methods for finding the PDF of the compound Poisson distribution $T_\nu$ for dwarf $\nu$. An early algorithm was developed by Panjer \cite{Panjer:1981aa} but we take advantage of a straightforward and efficient fast Fourier transform (FFT) method \cite{Embrechts:2009ly} which has also found use in astrophysics \cite{1957PCPS...53..764S,2009JCAP...07..007L,2010PhRvD..82l3511B}.

For a single dwarf $\nu$, let $\F_{w,\nu}$ be the Fourier transform (or characteristic function) of the probability distribution for the weight of a detected dark matter photon from $\nu$.  The Fourier transform of the PDF for $T_\nu$ (due to dark matter annihilation), denoted $\F_{{T_\nu}}$, is given by (see e.g. \cite{Embrechts:2009ly}),
\begin{equation}
\F_{{T_\nu}} = \exp[\mu_\nu\,(\F_{w,\nu} -1)].
\label{eqn:compound}
\end{equation}
To incorporate both signal and background photons into the PDF for $T$ we use the fact that a convolution is equivalent to multiplication in Fourier space. The full PDF for $T$ is
\begin{equation}
\F_T = \prod\limits_{\nu} \exp[\mu_{{\nu}} ( \F_{w,\nu} - 1)] \, \times \, \prod\limits_{\nu}\F_{B,\nu}
\label{eqn:Tpdf}
\end{equation}
where $\F_{B,\nu}$ is the Fourier transform of the empirically measured distribution of the sum of weights due to background processes for dwarf $\nu$.

In practice the Fourier transforms are performed using an FFT on a discrete grid of possible $T$ values. The single-event weight PDFs and the background PDFs are ``tilted'' \cite{Embrechts:2009ly} before taking the FFTs to form $\F_{w,\nu}$ and $\F_{B,\nu}$ and the PDF of $T$ is ``tilted back'' after applying the inverse FFT to $\F_T$. The tilting prevents aliasing which can be induced by the FFT.

The search for a line proceeds by first using Eqs.~\ref{eqn:Tpdf} \& \ref{eqn:wsearch} to derive the PDF of $T$ under the null hypothesis that there is no dark matter signal (all $\mu_\nu$'s are 0). The measured value of $T$, called $T^*$, is obtained by summing the weights of all photons in the 1$\degr$ ROIs centered on each of the dwarfs. The significance of the detection is the probability that $T$ would be measured to be less than $T^*$ if the null hypothesis were true. For example, if there is 99.7\% chance that $T<T^*$ then a line has been detected at 99.7\%, or $3\sigma$, significance.

The expected number of background counts $b_Q$ is found by fitting a power law to all photons within 15$\degr$ of the dwarf (excluding the central 1$\degr$ and the masked sources). For purposes of weighting, the background is assumed to be statistically isotropic, i.e. independent of angular separation from the ROI center. This may not be true in practice due to the presence of unresolved sources; however, the background sampling automatically includes any non-Poisson aspect of the background in the PDF of $T_\nu$ (or equivalently in $\F_{B,\nu}$).

The expected number of dark matter annihilation events $s_Q$, detected from a particular dwarf, with energy between $E$ and $E+dE$ and with angular separation in a solid angle interval $d\Omega(\theta)$ is
\begin{equation}
s_Q = J \, \frac{\sigv}{8\pi M_\chi^2}\frac{\dNg(E)}{dE} \, \eps(E) \, \PSF(E,\theta) \, dE \,d\Omega(\theta).
\label{eqn:sI}
\end{equation}
In the above $M_\chi$ is the mass of the dark matter particle, $\sigv$ is the velocity-averaged annihilation cross section into a pair of gamma-rays, and $dN_{\gamma}/dE$ is the number of photons per energy interval emitted per annihilation. The point spread function PSF$(E,\theta)$ is the probability per solid angle of detecting a photon of energy $E$ an angular distance $\theta$ from the source, and $\eps(E)$ is the detector exposure in units of cm$^{2}$s. The quantity $J$ quantifies the dark matter distribution within a particular dwarf \cite{2007PhRvD..75h3526S,2008ApJ...678..614S,2011MNRAS.418.1526C,2010PhRvD..82l3503E,2011ApJ...738...55M,2010ApJ...712..147A,2009JCAP...06..014M,2012arXiv1205.0311W,2011ApJ...733L..46W}.

We use the publicly available data from the Fermi Science Support Center (FSSC)\footnote{{\tt http://fermi.gsfc.nasa.gov/ssc/data/analysis/}} and version {\tt v9r27p1} of the Fermi Science Tools. We extract all photons of ${\tt evclass}=2$ using the tool ${\tt gtselect}$ in the Mission Elapsed Time interval [239557417 - 357485329] in the energy range between 8 GeV and 1 TeV, and with ${\tt zmax=100}$. We select good time intervals (with all standard recommendations as stated on the FSSC), and compute $\eps(E)$ and PSF$(E,\theta)$ using {\tt gtpsf} with the ${\tt P7SOURCE\_V6}$ instrument response functions. 

The dark matter annihilation is modeled as point source emission from each dwarf, and we utilize the values for $J$ given in \cite{2011PhRvL.107x1302A}.
For a line search the energy dispersion of the detector can be important. We incorporate this uncertainty by giving a width to $\dNg/dE$. The spectrum due to line emission is  simply $\dNg/dE = 2\delta(E-E_{\gamma})$ (but see also \cite{2012arXiv1205.1045T,2008JHEP...01..049B}). The 68\% containment on the energy uncertainty of Fermi is approximately 10\% for photon energies above 10 GeV. We model this energy uncertainty by setting the annihilation spectrum $\dNg/dE$ to be a Gaussian centered on $M_\chi$, normalized to 2, with a standard deviation of 10\% of the mean. We have reproduced the analysis with top hat distributions with widths from 5\% to 30\%. The effects are small and leave our conclusions unchanged.

\begin{figure}
\includegraphics[scale=0.85]{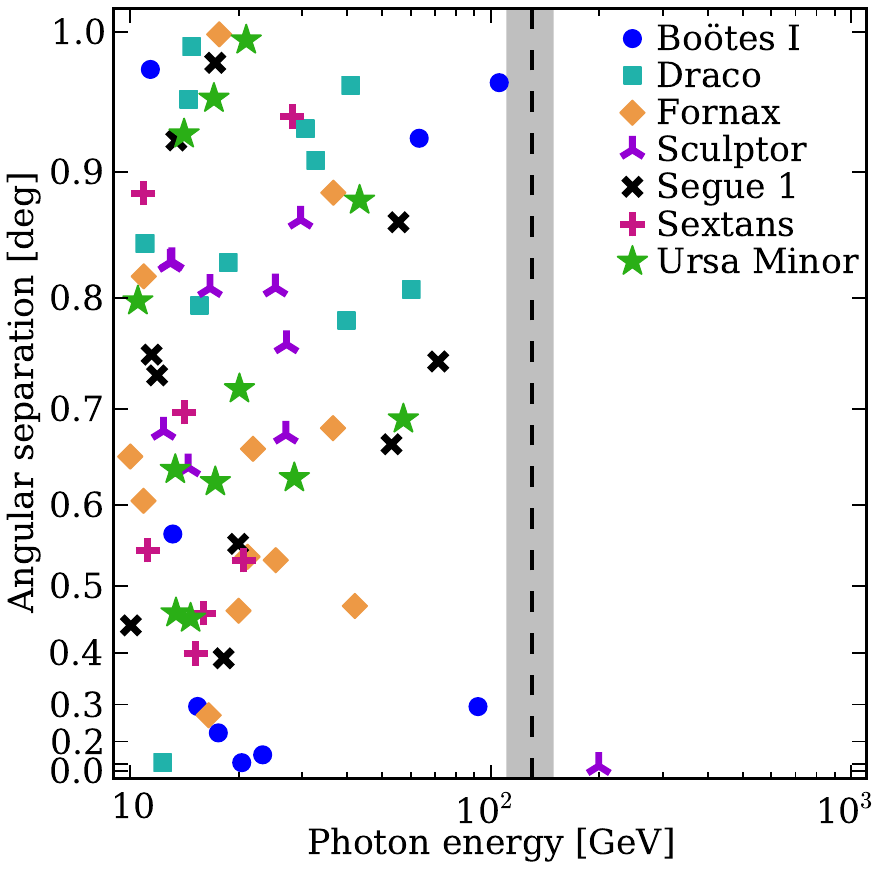}
\caption{\label{fig:photons}A snapshot of every photon having an energy between 10 GeV and 1 TeV that has been detected within 1$\degr$ of each of the seven dwarfs. The black dashed line is at 130 GeV \cite{2012arXiv1204.2797W,2012arXiv1205.1045T}. The gray region is $\pm15\%$ around 130 GeV, a rough gauge of the energy dispersion of the LAT. The vertical axis is scaled according to solid angle so that an isotropic distribution of photons will be spread uniformly along this axis.}
\end{figure}
%
%
Figure~\ref{fig:photons} shows the individual photon events between 10 GeV and 1 TeV that were detected within 1$\degr$ of each of the seven dwarfs. The vertical axis measures the angular separation between the event and the center of the dwarf. It is scaled according to solid angle so that an isotropic distribution of events should be distributed uniformly over the vertical axis. There are no photons with energy within 15\% of 130 GeV (gray shaded region). The 68\% energy resolution of the LAT ranges from about 8\% at 10 GeV to about 14\% at 1 TeV while the 68\% containment angle (PSF) varies from 0.3$\degr$ to $0.2\degr$ over this energy range (95\% containment is about 0.8$\degr$) \cite{LATperformance}.  We conclude that the {\it dwarfs show no evidence of a gamma-ray line at 130 GeV}.

\begin{figure}
\includegraphics[scale=0.85]{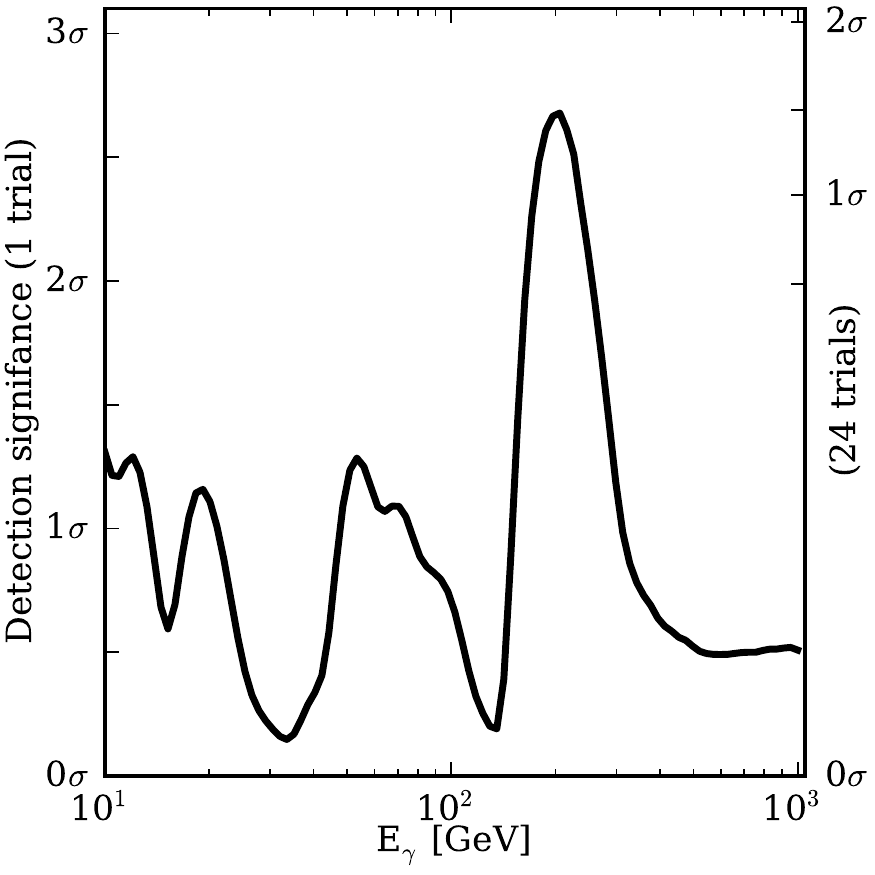}
\caption{\label{fig:search} Results of the a search for line emission using an optimized combined search of seven dwarf galaxies. The horizontal axis represents the energy of the gamma-ray line searched for. The left vertical axis is the significance of the detection (in terms of Gaussian standard deviations). The right vertical axis incorporates a trials factor of 24, roughly the number of independent energies searched. The non-significant peak at 200 GeV is due to a single photon from Sculptor (see Fig.~\ref{fig:photons}).}
\end{figure}

Using the formalism described above we perform a search for line emission over a range of energies. A Gaussian energy spectrum with a standard deviation of 10\% is used to calculate $s_Q$ (Eqs.~\ref{eqn:wsearch} and \ref{eqn:sI}). We perform a separate search for each possible line energy, taking 100 log-spaced steps from 10 GeV to 1 TeV. The results of the search are illustrated in Fig.~\ref{fig:search}.
%
%
%
Note that the inclusion of a trials factor dilutes the significance of any line. 
We can make a very rough estimate of the number of ``independent'' trials by assuming that a search for a line at $E_\gamma$ uses the photons in the window $E_\gamma(1\pm\alpha)$. If the energy of the $(-\alpha)$ edge of the window is $E_1$ the upper edge of the window is at an energy $E_2 = E_1 \, (1+\alpha)/(1-\alpha)$. Therefore the number of ``independent'' (i.e. non-overlapping) windows $n$ between $E_{\rm min}=8$ GeV and $E_{\rm max}=1$ TeV is specified by $E_{\rm max} = E_{\rm min} [ (1+\alpha )/( 1-\alpha)]^n$. 
An energy window of $\alpha=0.10$ corresponds to about 24 trials. On the right vertical axis of Fig.~\ref{fig:search} we plot the significance including a trials factor of 24 as a rough guide to the true significance of any tentative line. It is clear that the data do not strongly suggest that line emission is present at any energy.

Given that there is no evidence of line emission from the dwarfs we can place upper limits on the annihilation cross section into two photons. In this case, the weight choice analogous to Eq.~\ref{eqn:wsearch} that maximizes the signal to noise ratio is $w(Q) = s_Q / ( b_Q + s_Q)$ \cite{GSK12}. For each mass we find the cross section above which there is less than a 5\% chance of measuring the test statistic $T$ to be smaller than observed.
\begin{figure}
\includegraphics[scale=0.85]{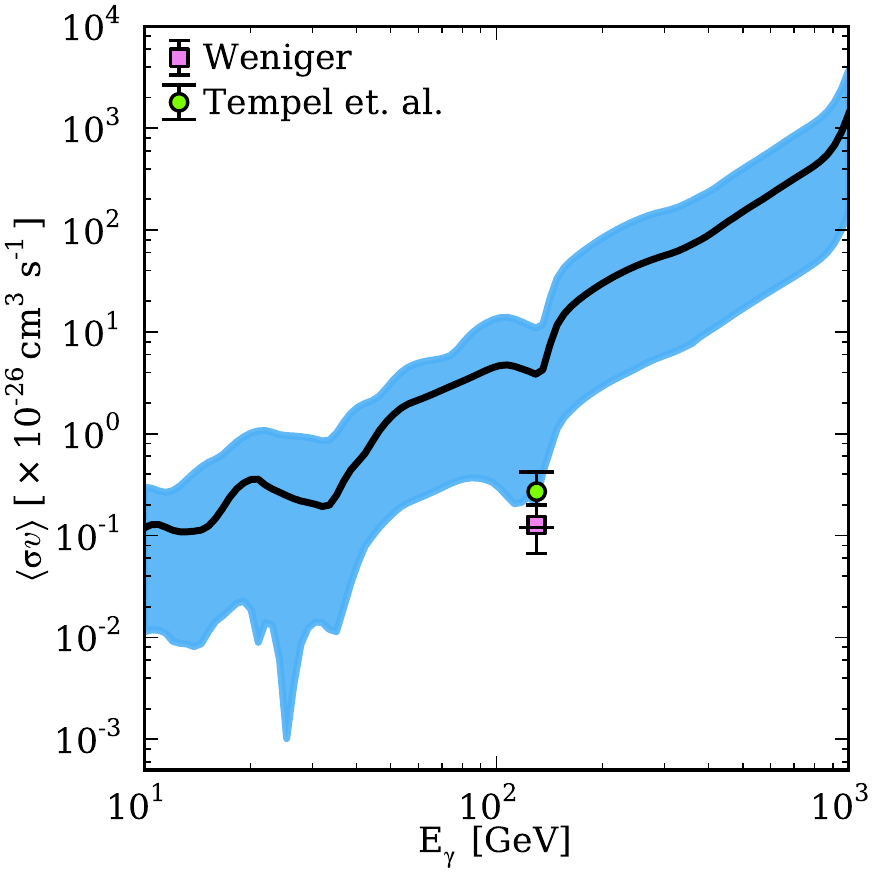}
\caption{\label{fig:limits}95\% upper limits on $\sigv$ for annihilation into a pair of photons each having energy $E_\gamma$. The black line is the limit using the best fit $J$ values for the dwarfs. The blue region corresponds to the 95\% systematic uncertainty in the estimates of $J$. The two points are the dark matter interpretations for the tentative signals observed by \cite{2012arXiv1204.2797W,2012arXiv1205.1045T} under the assumption of an Einasto dark matter profile and annihilation into two gamma-rays, with 95\% error bars.}
\end{figure}
%
%
The resulting upper limits are plotted in Fig.~\ref{fig:limits} (together with the results from \cite{2012arXiv1204.2797W,2012arXiv1205.1045T}). By far, the largest source of systematic uncertainty is in the $J$ values for the dwarfs. The black line in the figure is the limit found when the $J$ values are set to their best fit values found in \cite{2011PhRvL.107x1302A}. The effect of varying the $J$ values within their observational uncertainties is shown by the blue shaded region. One at a time, we set the $J$ value for each dwarf to its upper or lower 95\% error bar and recompute the 95\% cross section upper limit. The differences induced by each dwarf are added in quadrature to produce the boundaries of the shaded region. This procedure gives an estimate of the systematic effect due to the difficulty of determining each dwarf's dark matter distribution. 

For annihilation channels producing continuum emission (e.g. into heavy quark or lepton pairs) dwarf galaxies provide strong limits on the annihilation cross section \cite{2011PhRvL.107x1303G,2011PhRvL.107x1302A,2010ApJ...712..147A,2010PhRvD..82l3503E,2010JCAP...01..031S,2012arXiv1203.2954C,2012arXiv1203.6731M,2012arXiv1205.3620B,2012PhRvD..85f2001A,2010ApJ...720.1174A,2011JCAP...11..004R,2011APh....34..608H}. It is challenging to produce such limits from the Galactic center: despite the high dark matter density ($J$ value hundreds to thousands of times larger than the dwarfs) the astrophysical background cannot be easily subtracted or modeled. However, a gamma-ray line search is not hindered by these backgrounds to the degree that a continuum search is. For this reason, the  Galactic center may be a more attractive target when searching for line emission. The upper limits obtained by \cite{2012arXiv1203.1312B, 2012arXiv1204.2797W} are much stronger than those obtained here from the dwarf data. A recent search by the Fermi collaboration for gamma-ray lines in the  Galactic halo (including the Galactic center) \cite{2012arXiv1205.2739F} did not show evidence for a 130 GeV line and places stronger upper limits than found here.

It appears that the large increase in dark matter density, and the proximity of the Galactic center are much more constraining than are dwarf galaxies when it comes to line emission searches. At the present time dwarf galaxies can neither confirm nor deny a dark matter line interpretation of the Galactic center data.

SMK thanks the Texas Cosmology Center for hospitality and the organizers and participants of the "Dark Matter Signatures in the Gamma-ray Sky" workshop at the University of Texas-Austin for stimulating discussions that lead to this work.  AGS and SMK are supported by NSF grant PHY-0969853 and by a 2012 Salomon Award through Brown University.

\bibliography{manuscript}

\end{document}